\documentclass[journal,twoside,final]{IEEEtran}
\usepackage{epsf,amsmath,amssymb,amsfonts,times,bm,xcolor}
\usepackage{epsf,epsfig,graphicx}
\usepackage{enumerate}
\usepackage{srcltx}
\DeclareMathAlphabet{\mathitbf}{OML}{cmm}{b}{it}
\usepackage{psfrag}
\usepackage{cite,multirow}
\usepackage{algorithmic}
\usepackage{algorithm,setspace}
\usepackage{framed}
\usepackage{stfloats}
\usepackage{textcomp}
\usepackage{balance,cite}
\usepackage{subcaption}
\usepackage{bbm}

\def \min{\operatornamewithlimits{min}}
\def \max{\operatornamewithlimits{max}}

\def \log{\operatorname{log}}

\def \beqi{\begin{IEEEeqnarray}{rcl}\IEEEyesnumber}
\def \eeqi{\end{IEEEeqnarray}}

\def \beq  { \begin{equation} }
\def \eeq { \end{equation} }
\def \beqn{ \begin{eqnarray} }
\def \eeqn{ \end{eqnarray} }
\def \bmat{\begin{bmatrix}}
\def \emat{\end{bmatrix}}
\def \bmats{\left[\begin{smallmatrix}}
\def \emats{\end{smallmatrix}\right]}

\newtheorem{definition}{Definition}

\newtheorem{remark}{Remark}

\def \log{\operatorname{log}}

\def \be {\begin{eqnarray}}
\def \ee {\end{eqnarray}}
\def \ben {\begin{eqnarray*}}
\def \een {\end{eqnarray*}}

\newcommand{\N}{\mathbbmss{N}}
\newcommand{\R}{\mathbbmss{R}}

\begin{document}
\title{\textsf{DIR-ST$^2$}: Delineation of Imprecise Regions Using Spatio--Temporal--Textual Information}
\author{Cong Tran,~Won-Yong Shin,~\IEEEmembership{Senior Member,~IEEE}, and Sang-Il Choi,~\IEEEmembership{Member,~IEEE}
\thanks{The present research was conducted by the research fund by Dankook University in 2016 and was supported by the Basic Science Research Program through the National Research Foundation of Korea (NRF) funded by the Ministry of Education (2017R1D1A1A09000835) and by the National Research Foundation of Korea Grant through the Korean Government (MSIT) under Grant 2018R1A2B6001400. 
\newline {\em (Co-corresponding authors: Won-Yong Shin and Sang-Il Choi.)}}
\IEEEcompsocitemizethanks{\IEEEcompsocthanksitem The authors are with the Department of Computer Science and Engineering, Dankook University, Yongin 16890, Republic of Korea \protect (e-mail: trancong208@gmail.com; wyshin@dankook.ac.kr; choisi@dankook.ac.kr).}
}
%
\maketitle



\IEEEpubidadjcol

\begin{abstract}
An imprecise region is referred to as a geographical area without a clearly-defined boundary in the literature. Previous clustering-based approaches exploit spatial information to find such regions. However, the prior studies suffer from the following two problems: the subjectivity in selecting clustering parameters and the inclusion of a large portion of the undesirable region (i.e., a large number of noise points). To overcome these problems, we present \textsf{DIR-ST$^2$}, a novel framework for delineating an imprecise region by iteratively performing {\em density-based clustering}, namely DBSCAN, along with not only {\em spatio--textual} information but also {\em temporal} information on social media. Specifically, we aim at finding a proper radius of a circle used in the iterative DBSCAN process by gradually reducing the radius for each iteration in which the temporal information acquired from all resulting clusters are leveraged. Then, we propose an efficient and automated algorithm delineating the imprecise region via hierarchical clustering. Experiment results show that by virtue of the significant noise reduction in the region, our \textsf{DIR-ST$^2$} method outperforms the state-of-the-art approach employing one-class support vector machine in terms of the $\mathcal{F}_1$ score from comparison with precisely-defined regions regarded as a ground truth, and returns apparently better delineation of imprecise regions. The computational complexity of \textsf{DIR-ST$^2$} is also analytically and numerically shown.
\end{abstract}


\section{Introduction}\label{intro}
\subsection{Background}\label{background}

An imprecise region (also known as a vague or vernacular region) is referred to as a geographical area with (administratively) nonexistent boundaries or no clearly-defined boundary in the literature, which reveals that such a region cannot be objectively visualized and is dependent solely upon a personal point of view. There are a variety of examples of imprecise regions around the world, such as the South of the US and the Midlands of the UK. The problems of uncertainty and approximation in geographic information retrieval were discussed in~\cite{geographic}, which shows that due to the lack of boundary information for imprecise regions, applications of information retrieval and spatial browsing are incapable of appropriately searching for items or locations inside the regions. For example, query processing for famous restaurants located in the Midlands of the UK may not be performed successfully or cost-effectively. Therefore, it is of significant importance both academically and commercially to delineate imprecise regions with their geographic boundaries.

On the other hand, online social media can be thought of as useful sources of information for solving the delineation problem of imprecise regions. Supporting geo-referenced functions (e.g., geo-tagged tweets on Twitter), popular real-world social media such as Twitter~\cite{twitter1,twitter2}, Facebook~\cite{facebook}, Flickr~\cite{flickr}, and Foursquare~\cite{4square} enable users to describe that their visiting locations are related to a region of interest by checking in online or posting photos of their visit. By virtue of the collected data from social media, an imprecise region can be discovered by performing density-based clustering based on the geo-tagged records in which name tags of a region of interest are contained~\cite{rw4,rw9,rw10}.

However, previous studies on the imprecise region delineation adopting density-based clustering methods~\cite{rw4,rw9,rw10} suffer from two fundamental problems. First, the selection of clustering parameters is often subjective since it depends on the distribution of geo-tagged points in a given dataset~\cite{optics}. Second, the delineation accuracy may not be satisfactory due to the existence of noise points that are dense enough to form clusters, e.g., a large number of geo-tagged records that are textually relevant to a region of interest but were generated in popular metropolitan cities. Such noise points act as a severe obstacle for prior approaches to efficiently discover the footprints of imprecise regions. For example, a large portion of some metropolitan areas can often be included in the desired regions, which leads to the delineation performance degradation. 

Since two-dimensional spatial information (i.e., latitude--longitude pairs of users) is insufficient to solve the aforementioned problems, additional information such as the terrain elevation of a region of interest was taken into account to enhance the delineation accuracy in~\cite{rw8}. However, the temporal information (i.e., a timestamp), which can also be acquired from one field of collected social media data, has not yet been unveiled in delineating imprecise regions. 

\subsection{Motivation and Main Contributions}\label{contrib}

In this paper, we present \textsf{DIR-ST$^2$}, a novel framework for delineating an imprecise region exploiting the {\em temporal} information as well as the {\em spatio--textual} information on geo-located Twitter. A key component of the proposed framework is to {\em iteratively} perform density-based spatial clustering of applications with noise (DBSCAN)~\cite{dbscan}, which is the most commonly used density-based clustering algorithm. Specifically, we aim at finding a proper input parameter $\varepsilon$ in the iterative DBSCAN process by gradually reducing $\varepsilon$ for each iteration, where $\varepsilon$ represents the maximum radius of the neighborhood from a point in all clusters. 

Unlike the prior studies in~\cite{rw4,rw9,rw10}, our delineation problem is formulated by observing resulting clusters via DBSCAN using the temporal information of tweets---people are likely to {\em regularly} mention the area in which they reside during their daily activities while mentioning other regions randomly and intermittently. Based on the observations, we determine a stopping criterion of the iterative DBSCAN process. More specifically, the parameter $\varepsilon$ is decreased by a small constant for each iteration until the temporal regularity condition, expressed as the Shannon entropy, of tweets in the major cluster is not fulfilled, where the major cluster corresponds to the one having the largest number of points among clusters. The imprecise region is finally delineated from the major cluster found at the second last iteration. In addition, we present an idea of incorporating an efficient and automated algorithm via hierarchical clustering into our \textsf{DIR-ST$^2$} method. In the algorithm, instead of brute-force search over $\varepsilon>0$, we investigate the values of $\varepsilon$ that lead only to different clustering results obtained by DBSCAN, thus resulting in the reduced complexity. Experimental results demonstrate that our \textsf{DIR-ST$^2$} framework shows superior performance to the state-of-the-art approach delineating imprecise regions.

Our main technical contributions are summarized as follows:
\begin{itemize}
\item design of a novel framework, named \textsf{DIR-ST$^2$}, that delineates an imprecise region by iteratively performing DBSCAN along with the spatio--temporal--textual information on social media;
\item incorporation of an efficient and automated algorithm that delineates the imprecise region via hierarchical clustering into the \textsf{DIR-ST$^2$} framework;
\item validation of our \textsf{DIR-ST$^2$} approach through intensive experiments by showing the superiority over the state-of-the-art method employing the one-class support vector machine (OCSVM) algorithm~\cite{rw6} with both precisely-defined regions and imprecise regions;
\item analysis and numerical evaluation of the computational complexity.
\end{itemize}

Our framework sheds light on a better understanding of how to intelligently exploit the spatio--temporal--textual information for more accurate imprecise region delineation.

\subsection{Organization}\label{organization}
The rest of the paper is organized as follows. In Section~\ref{relatedwork}, we summarize previous studies that are related to our work. Section~\ref{data} describes our data acquisition and processing steps. The overall methodology of \textsf{DIR-ST$^2$} is presented in Section~\ref{methodology}. Implementation details of the proposed framework with an efficient and automated clustering algorithm are shown in Section~\ref{algorithm}. Experimental results are provided in Section~\ref{experiment}. In Section~\ref{conclusion}, we summarize the paper with some concluding remarks.
\subsection{Notation}\label{notation}
Throughout this paper, all logarithms are assumed to be to the base 10. Table~\ref{tab:notation} summarizes the notations used in this paper, which will be formally defined in the following sections when we introduce our problem formulation and technical details.
\begin{table}[]
\centering
\caption{Summary of notations}
\label{tab:notation}
\begin{tabular}{p{1.2cm}| p{6.8cm}}
\hline
Notation               & Description                                 \\ \hline
$\mathcal{X}$                    & set of geo-tagged points                            \\
$n$                    & number of geo-tagged points                            \\
$\varepsilon$                    & radius of a circle in the clustering process                           \\
$MinPts$                    & minimum number of points in an $\varepsilon$-neighborhood                            \\
$\mathcal{C}$                    & set of clusters                            \\
$N$ & number of clusters                            \\
$noise$ & set of noise points					\\
$T_{i,j}$ &$j$th tweeted time in cluster $C_i$					\\
$I_{i,j}$ & inter-tweet time interval between $j$th and $(j + 1)$th tweeted times				\\
$H(I_{i,j})$ & Shannon entropy of $I_{i,j}$				\\
$\delta$ & threshold associated with the temporal regularity condition					\\
$Prec$ & precision					\\
$Rec$ & recall					\\
$\mathcal{F}_1$ &  $\mathcal{F}_1$ score \\
\hline
\end{tabular}
\end{table}
\section{Related Work}\label{relatedwork}

In this section, we briefly summarize the prior work in three areas of research that are closely related to our topic, namely spatial clustering, imprecise region delineation, and the combination of these two areas. 

\textbf{Spatial clustering}. DBSCAN~\cite{dbscan} has been known as one of the most popular spatial clustering algorithms~\cite{ieacessreview,ieeeaccess2} due to the capability of indexing separated clusters, the robustness in detecting outliers, and the ability to provide arbitrarily-shaped clusters. Because of its popularity, many variants of DBSCAN have been extensively developed in different applications: a generalized version of DBSCAN that differently measures the cardinality of the neighborhood of a point was proposed in~\cite{gdbscan}; another DBSCAN algorithm was presented in~\cite{stdbscan} by generating clusters based on both the spatial and temporal attributes; hierarchical DBSCAN (HDBSCAN) was introduced in~\cite{hdbscan} as an automated method that measures the stability of clusters in a clustering hierarchy to select proper input parameters of DBSCAN; density-based spatio--textual clustering (DBSTexC) was introduced in~\cite{minhasonam} to improve the clustering accuracy by leveraging both relevant and irrelevant geo-tagged points to a region of interest; and an effective density-based clustering framework (DCF) was presented in~\cite{dcf} by integrating a neighborhood density estimation model into the underlying DBSCAN framework.

\textbf{Imprecise region delineation}. To delineate an imprecise region, researchers have acquired data either from human opinions~\cite{rw1,rw2} or from computer applications~\cite{rw3,rw4,rw7,rw9,rw10}. With the help of volunteers, an empirical study was conducted in~\cite{rw1} by generating a probabilistic representation of the Downtown Santa Barbara area. Since the task of collecting such empirical data is often expensive and thus is not viable in a wide range of real applications, computer-based approaches have been developed as an alternative. The method in~\cite{rw3} exploited the textual content on the web to discover the desired imprecise region. Another similar approach made use of online yellow pages to geo-locate business names inside a vernacular region, thus unfolding the region's location~\cite{rw7}. With the rapid growth of online social media, geo-tagged points have been widely used for analyzing regions of interest, including imprecise regions~\cite{rw4,rw9,rw10}. However, since the collected data may be textually heterogeneous, they inevitably contain noise in this context.

\textbf{Imprecise region delineation using spatial clustering}. To enhance the accuracy of imprecise region delineation, it is straightforward to adopt spatial clustering methods to precisely deal with noise points. One of the baseline methods based on kernel density estimation (KDE) was developed in~\cite{rw5} by creating a fuzzy boundary of an imprecise region from the densest area of points. In~\cite{rw6}, an automated method based on OCSVM was introduced by generating a crisp boundary of imprecise regions. The input parameter of OCSVM was obtained by performing the statistical analysis on regions with administrative boundaries in Europe. Later, the authors in~\cite{rw8} extended the method in~\cite{rw6} by incorporating more training data, which include place semantics from Flickr photo tags, population counts, terrain elevation, and land coverage information. The methods in~\cite{rw5,rw6,rw8} showed satisfactory performance along with well-defined geographic boundaries (i.e., administrative boundaries). Recently, an algorithm with linear complexity in the number of geo-tagged points was presented in~\cite{dungpaper} by estimating the boundary of a region of interest in the form of a circle.

\section{Data Acquisition and Processing}\label{data}
For data acquisition, we use Twitter Streaming Application Programming Interface (API)~\cite{twitter}. Our dataset is composed of a large set of geo-tagged tweets collected from Twitter users for two months from June 1, 2016 to July 30, 2016 in the UK.\footnote{Unlike most studies on the Twitter network~\cite{minhasonam,bot1}, we do not remove the tweets that were likely to be generated by automated services (e.g., tweetbots), because such tweets rather play a crucial role in delineating imprecise regions more accurately. This comes from the fact that tweetbots driven by timers tend to exhibit a regular behavior~\cite{botdetect}, and the regularity over time is leveraged by our method.} We observe that each tweet contains a number of entities that can be differentiated by their attributed field names. To extract the textual, geographical, and temporal information from the collected dataset, we adopt four essential fields as follows:
\begin{itemize}
    \item   \emph{text}: actual UTF-8 text of the status update;
    \item   \emph{lat}: latitude of the tweet's location, measured in degree;
    \item   \emph{lon}: longitude of the tweet's location, measured in degree;
    \item   \emph{created\_at}: the GMT time when the tweet is created.
\end{itemize}

Note that the two location fields, \emph{lat} and \emph{lon}, correspond to spatial (geo-tagged) information while the last field, \emph{created\_at}, represents temporal (time-stamped) information. Since the tweets generated at night time are insignificant, we select only the tweets created between 8am and 8pm in the BST time setting to maintain the continuity of temporal information. Due to the fact that Twitter users have a tendency to tag or to mention the name of a geographic region in their tweets to express their interest in the region, we can easily query all region-relevant tweets by searching for keywords associated with the region in the users' {\em text} field. Since users may misspell the name of a region or mention it using different names in their tweets, our query processing includes the search based on keywords that are {\em semantically coherent} with the name of a geographic region such as its abbreviated names, its nicknames, etc. (e.g., London and Londinium).

\begin{figure*}[t]
\centering 
\begin{subfigure}[]{0.85\textwidth}
\centering 
\includegraphics[width=6.5in]{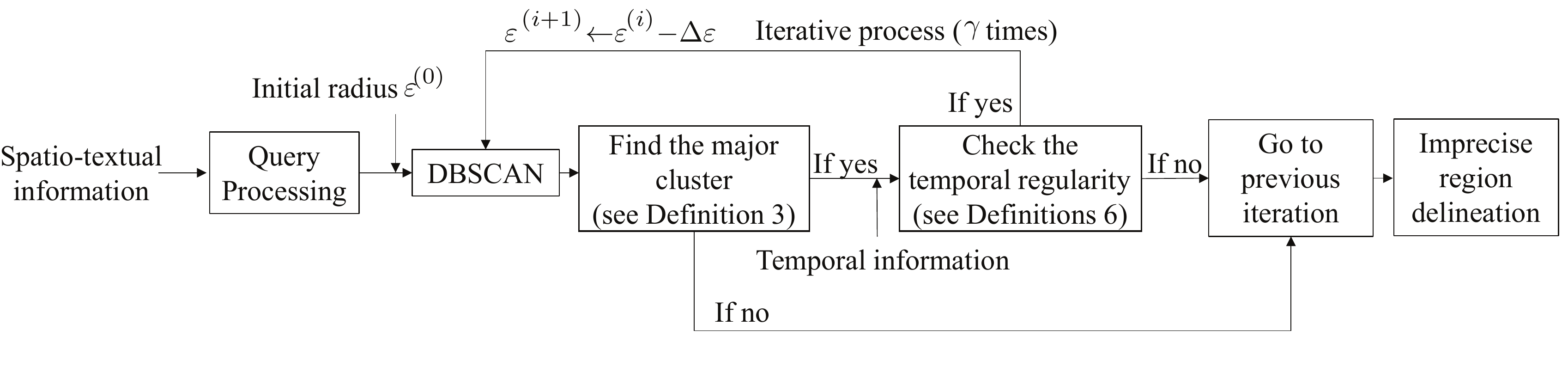}
\caption{The overall steps of \textsf{DIR-ST$^2$}.}
\label{fig:overall1}
\end{subfigure}
\begin{subfigure}[]{0.85\textwidth}
\centering 
\includegraphics[width=5.5in]{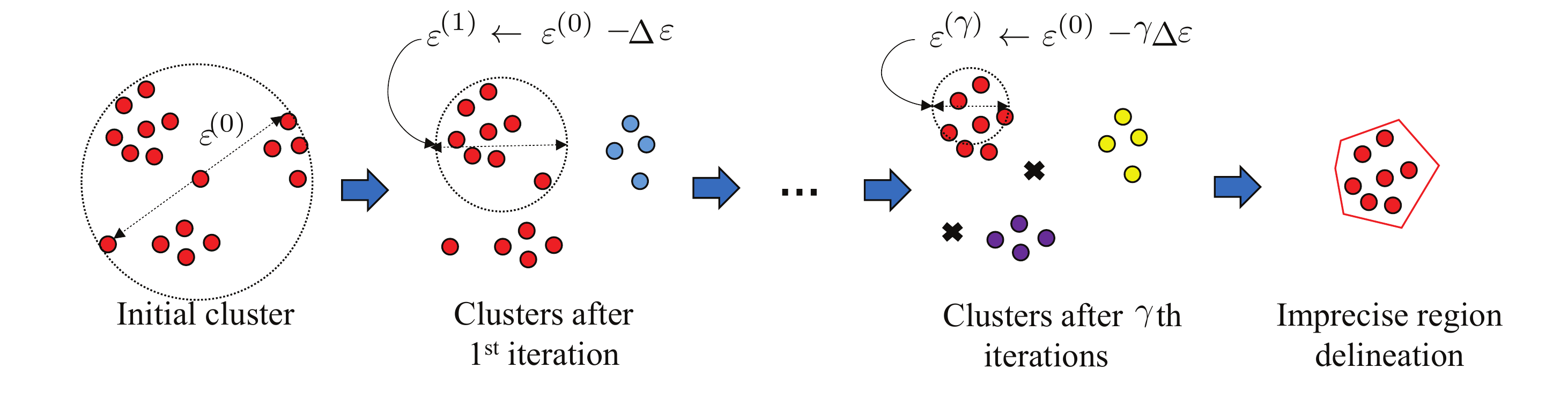}
\caption{Illustration of clusters for each iteration, where each cluster is depicted by circles of the same color and noise points are depicted by black crosses.}
\label{fig:overall2}
\end{subfigure}
\caption{The schematic overview of our \textsf{DIR-ST$^2$} approach.}
\label{fig:overall}
\end{figure*}
\section{Methodology}\label{methodology}
In this section, we present the overview of our \textsf{DIR-ST$^2$} framework using geo-tagged tweets, where the overall procedure and a brief introduction to DBSCAN are shown. Then, motivated by the observations for resulting clusters, we then formulate our clustering problem in terms of finding a proper input in the DBSCAN stage using the temporal information.
\subsection{The Overall Procedure}\label{sec:overall}
In this subsection, we briefly describe our approach to delineating an imprecise region. Using the spatio--textual information from the {\em text}, {\em lat}, and {\em lon} fields in the collected dataset, we first query  the name of a certain imprecise region and its semantically coherent variations to acquire geo-tagged points relevant to the region of interest. Then, we perform DBSCAN {\em iteratively} by changing the value of $\varepsilon^{(i)}$ for the $i$th iteration, where $\varepsilon^{(i)}$ is a crucial input parameter of the DBSCAN algorithm and denotes the radius of a circle used in the clustering process. Now, let us explain a stopping criterion of this iterative process.
Using the temporal information extracted from the {\em created\_at} field, the parameter $\varepsilon^{(i)}$ is decreased by a small constant $\Delta\varepsilon>0$ for every iteration until the {\em temporal regularity condition} of tweets in the {\em major} cluster is not fulfilled (see Definition~\ref{def:6} for the temporal regularity condition). Here, the major cluster is referred to as the cluster such that the number of points in the cluster is higher than that of other clusters and the so-called noise points (see Definition~\ref{def:1} for more details). Finally, we are capable of delineating the imprecise region from the major cluster found at the second last iteration. The overall procedure is summarized in Fig.~\ref{fig:overall}, where the number of iterations is assumed to be $\gamma$.

\subsection{Density-based Clustering}\label{sec:dbscan}
In this subsection, we describe how to perform the original DBSCAN algorithm~\cite{dbscan}, which is the most commonly used density-based clustering. For the $i$th iteration of our \textsf{DIR-ST$^2$} method, DBSCAN operates based on two input parameters $\varepsilon^{(i)}$ and ${MinPts}$.\footnote{To simplify notations, $\varepsilon^{(i)}$ will be written as $\varepsilon$ if dropping the superscript $(i)$ does not cause any confusion.} Let $\mathcal{X}=\{x_1,x_2,\cdots,x_n\}$ denote the set of $n$ geo-tagged points, where each point $x_i$ represents the coordinate consisting of the latitude and longitude. In our method, we deal conveniently with the straight-line Euclidean distance $d(x,y)$ between any points $x,y \in \mathcal{X}$.\footnote{The shortest path between two locations measured along the surface of the Earth can also be taken into account by assuming that the Earth is spherical.} We begin by presenting the following two important definitions.
\begin{definition}\label{def:neighbor}
Given a parameter $\varepsilon \in R^+$, the {\em neighborhood set} for a point $x \in \mathcal{X}$ is denoted by $\mathcal{N}(x;\varepsilon)$ and is defined as 
\begin{equation}
\mathcal{N}(x;\varepsilon) = \{y \in \mathcal{X}|d(x,y) \leq \varepsilon\}.\nonumber
\end{equation}
\end{definition}
\begin{definition}\label{def:corepoint}
A point $x \in \mathcal{X}$ with at least $MinPts \in \R^+$ points around its $\varepsilon$-neighborhood is defined as a {\em core point}, i.e., $|\mathcal{N}(x;\varepsilon)| \geq MinPts$, where $|\mathcal{N}(x;\varepsilon)|$ is the cardinality of the set $\mathcal{N}(x;\varepsilon)$.
\end{definition}

From the above definitions, it is seen that $\varepsilon$ is the maximum radius of the neighborhood from a point and $MinPts$ is the minimum number of points required to form a dense region. All the points within the $\varepsilon$-neighborhood of a core point are classified into the same cluster, and points that do not belong to any clusters are labeled as {\em noise}. The overall steps of the DBSCAN algorithm are briefly summarized as follows:
\begin{itemize}
\item {\bf Step 1:} Specify $\varepsilon$ and $MinPts$. 
\item {\bf Step 2:} Mark all points in the set $\mathcal{X}$ as unclassified. 
\item {\bf Step 3:} Find all unclassified core points and assign points to clusters.
\item {\bf Step 4:} If no more core points are found, then label all clusters and noise points.
\end{itemize}

\begin{figure}
\centering 
\includegraphics[height=2in]{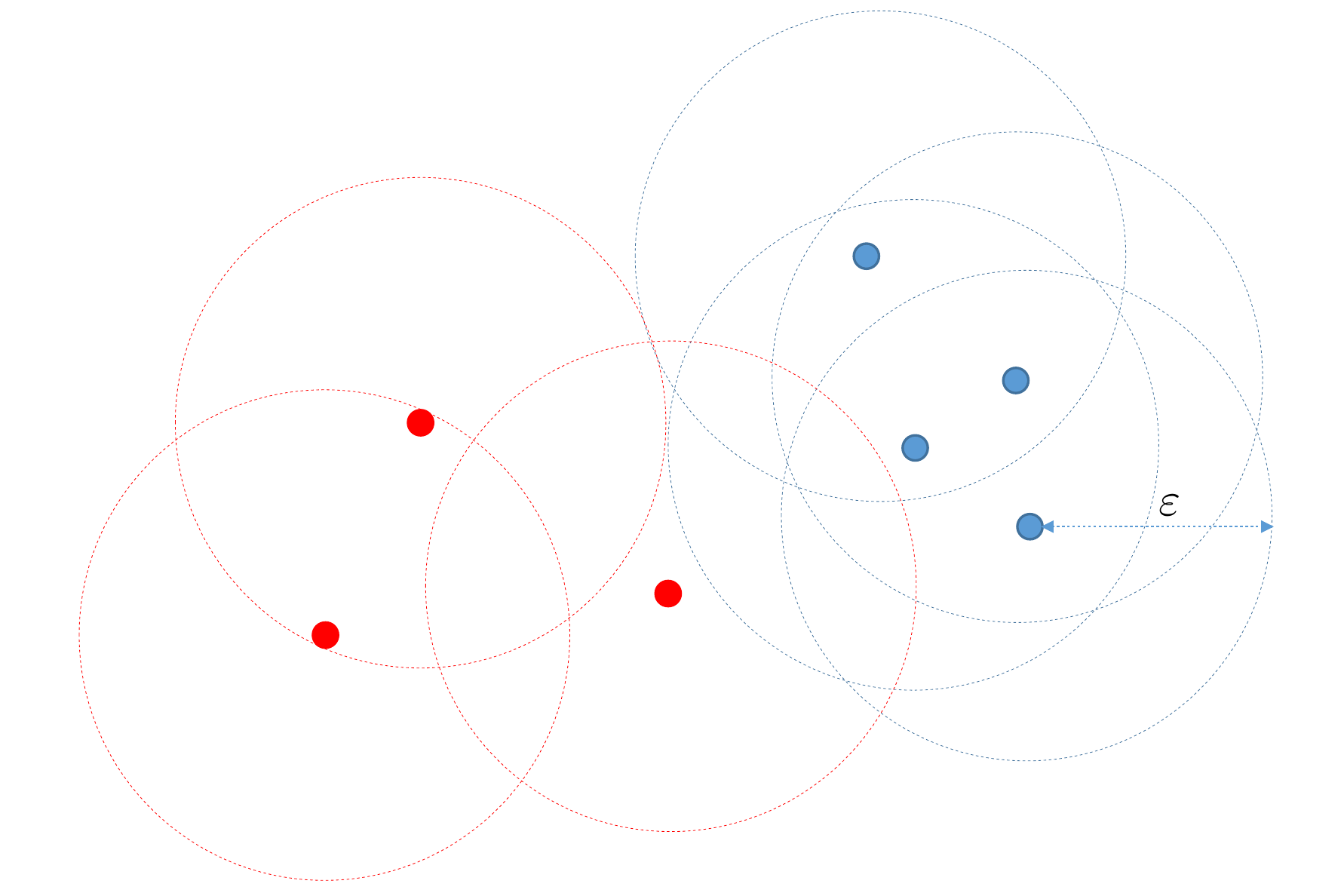}
\caption{An example of the resulting cluster via DBSCAN, where points in the cluster and noise points are depicted with blue and red colors, respectively, and the parameter $MinPts$ is set to 3.}
\label{fig:dbscan}
\end{figure}

As an example, the resulting cluster of DBSCAN is illustrated in Fig.~\ref{fig:dbscan}, where $MinPts$ is set to 3. Since the input parameters $\varepsilon$ and $MinPts$ remarkably affect clustering performance on the accuracy, it is important how to determine $MinPts$ in our \textsf{DIR-ST$^2$} framework.
\begin{remark}
The value of $MinPts$ can be decided adaptively based on both $\varepsilon$ and attributes of geo-tagged points. Suppose that $P_i$ denotes the number of {\em distinct} geo-tagged points within the $\varepsilon$-neighborhood of point $x_i\in \mathcal{X}$ (i.e., geo-tags superimposed at one geo-location are treated as one geo-tag when we count the number of distinct points). Then, as in~\cite{minpts}, we set the value of $MinPts$ to the expectation of the number of distinct points, $P_i$, over $i\in\{1,\cdots,n\}$, i.e.,
\begin{equation}\label{eq:minpts}
MinPts = \frac{1}{n}\displaystyle\sum_{i=1}^{n} P_i. 
\end{equation}
This parameter setting enables us to form clusters in dense regions. Note that since $P_i$ for $i\in \{1,\cdots,n\}$ is expressed as a function of $\varepsilon$, decision on $\varepsilon$ would automatically lead to a value of $MinPts$ accordingly from~(\ref{eq:minpts}). We refer to Appendix A for more details of this adaptive parameter setting.
\end{remark}
\begin{figure}
\centering 
\includegraphics[width=2.6in]{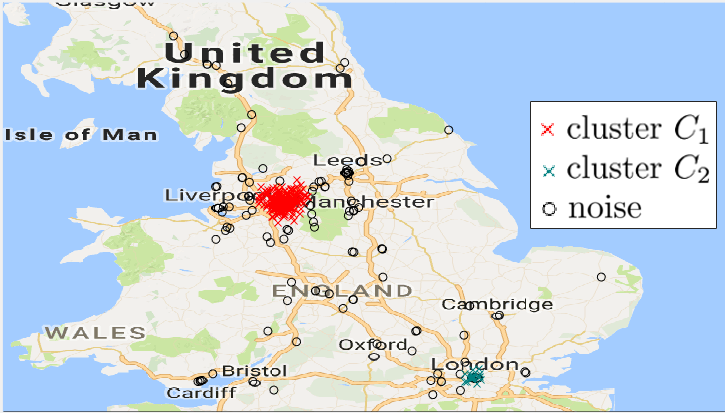}
\caption{The result of DBSCAN for Manchester with a proper setting of input parameter $\varepsilon$.}
\label{fig:manc}
\end{figure}
\begin{figure*}[ht]
\centering 
\begin{subfigure}[]{0.85\textwidth}
\centering 
\includegraphics[width=6.5in]{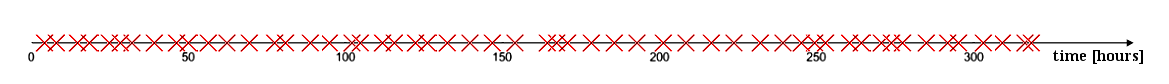}
\caption{Cluster in Manchester}
\end{subfigure}
\begin{subfigure}[]{0.85\textwidth}
\centering 
\includegraphics[width=6.5in]{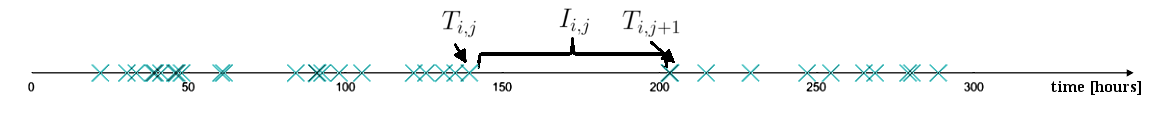}
\caption{Cluster in London}
\end{subfigure}
\caption{Tweeted time [hours]. Here, each cross represents the time when a tweet is posted for one month, starting from June 1, 2016.}
\label{fig:entropy}
\end{figure*}
\begin{figure}
\centering 
\includegraphics[width=2.2in]{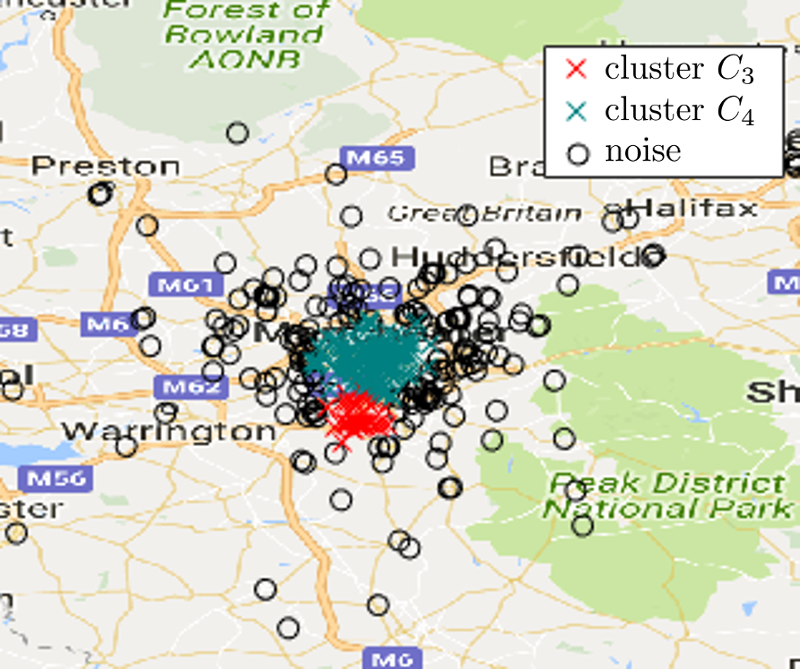}
\caption{The result of DBSCAN for Manchester when $\varepsilon$ is too small.}
\label{fig:eps2}
\end{figure}
\subsection{Observations and Problem Formulation}\label{problemdef}

In this subsection, we elaborate on the formulation of our clustering problem based on the the observations for resulting clusters using the temporal information of tweets. We first denote $\mathcal{C} = \{C_1,\cdots,C_{N}\}$ as the set of clusters returned by the DBSCAN algorithm, $N$ as the number of clusters, $\mathcal{X}_{C_i}$ as the set of geo-tagged points in cluster $C_i$ for $i \in \{1,\cdots,N\}$, and $noise$ as the set of noise points. Then, the {\em major} cluster is formally defined as follows. 

\begin{definition}\label{def:1}
Let $|\mathcal{X}_{C_i}|$ be the cardinality of a set $\mathcal{X}_{C_i}$ for $i \in \{1,\cdots,N\}$, and $|noise|$ be the cardinality of $noise$, respectively. Then, a cluster $C_m \in \mathcal{C}$ is defined as the {\em major} cluster if $|\mathcal{X}_{C_m}| > |\mathcal{X}_{C_i}|$ and $|\mathcal{X}_{C_m}| > |noise|$ for all ${i} \ne {m}$.
\end{definition}

As mentioned in Section~\ref{sec:overall}, DBSCAN is performed iteratively by gradually reducing $\varepsilon$ for each iteration, where $\varepsilon$ is initially set to a certain value so that all geo-tagged points in the dataset are covered by a single cluster and $MinPts$ is given in~(\ref{eq:minpts}). Figure~\ref{fig:manc} illustrates the result of the DBSCAN algorithm for Manchester when $\varepsilon$ is manually set to such an appropriate value that the major cluster (depicted by red crosses) fits into the administrative boundary of the Greater Manchester. It is seen that the region obtained from the major cluster can be accurately delineated. Then, a natural question is how to find such a proper clustering parameter. 

To answer this question, we start by introducing the following two crucial definitions. 

\begin{definition}\label{def:2}
Given that tweets are sorted in chronological order, let $T_{i,j}$ be the $j$th tweeted time in cluster $C_i$, where $i \in \{1,\cdots,N\}$ and $j \in \{1,\cdots,|\mathcal{X}_{C_i}|\}$. Then, the {\em inter-tweet time interval} between the $j$th and $(j+1)$th tweeted times, denoted by $I_{i,j}$, is computed by $I_{i,j}=T_{i,j+1}-T_{i,j}$. 
\end{definition}
\begin{definition}\label{def:3}
The Shannon entropy $H(I_i)$ of the variable $I_{i,j}$ for all $j\in\{1,\cdots,|\mathcal{X}_{C_i}| -1\}$ is defined as~\cite{entropy}
\begin{equation}\label{eq:entropy}
H(I_i) = -\sum_{j=1}^{|\mathcal{X}_{C_i}|-1}\mathbb{P}(I_{i,j})\log\mathbb{P}(I_{i,j}), 
\end{equation}
\end{definition}
where $i\in\{1,\cdots,N\}$.

Note that the entropy of $I_{i,j}$ measures the {\em regularity} of the set of inter-tweet time intervals. Let us recall the result of DBSCAN for Manchester shown in Fig.~\ref{fig:manc}. In this example, tweeted times [hours] are depicted in Fig.~\ref{fig:entropy}, where the tweets in two clusters $C_1$ and $C_2$, belonging to the Greater Manchester and London metropolitan areas, respectively, are used. From the figure, it is seen that the tweets in the desired cluster (i.e., the major cluster) tend to occur more regularly in time than those in another (undesired) cluster. This is because people are likely to {\em regularly} mention the area in which they reside during their daily activities while mentioning other regions randomly and intermittently. From~(\ref{eq:entropy}), it is shown that the entropy of inter-tweet time intervals for the clusters in Manchester and London is 1.31 and 4.36, which implies that tweets in Manchester tend to appear more regularly in time. It is also worth noting that the tweets in other clusters leading to a high entropy should also be treated as noise since they are not located in Manchester.

Next, let us turn to addressing the case where $\varepsilon$ is too small. Figure~\ref{fig:eps2} demonstrates the result of DBSCAN when $\varepsilon$ is too small (i.e., smaller than the value that we set in Fig.~\ref{fig:manc}). As illustrated in the figure, the major cluster $C_1$ in Fig.~\ref{fig:manc} is divided into two smaller clusters $C_3$ and $C_4$. From the fact that cluster $C_4$ is selected as the major cluster, a portion of Manchester represented by cluster $C_3$ would be inadvertently discarded in the region delineation. This is undesirable because cluster $C_3$ should not be regarded as noise. Interestingly, we also observe that the entropy values obtained from these two clusters $C_3$ and $C_4$ are both low and are similar to each other. This is due to the fact that the tweets in both clusters are still posted by people living in Manchester and thus contain the name of the region in the {\em text} field regularly. 

When DBSCAN is performed iteratively for other geographic regions, the above observations can also be made in a similar manner. Thus, for an accurate delineation of regions, it would be of significant importance to carefully choose the value of $\varepsilon$ not to be too small or too large. To this end, we formally define the {\em temporal regularity condition} as follows.
\begin{definition}\label{def:6}
For a given major cluster $C_m$, the temporal regularity condition is fulfilled if
\begin{equation}\label{eq:regularity}
H(I_i)-H(I_m) \geq \delta
\end{equation}
for all other clusters $C_i\neq C_m$, where $H(I_i)$ and $H(I_m)$ are the entropy of the variables $I_{i,j}$ and $I_{m,j}$ for all $j$, respectively, and $\delta>0$ denotes a pre-defined threshold.
\end{definition}

We finally aim at finding a proper $\varepsilon$ by reducing $\varepsilon$ by a small constant $\Delta\varepsilon>0$ for each iteration, while DBSCAN is performed iteratively by checking the temporal regularity condition with the major cluster. The major cluster found at the second last iteration is chosen to delineate the imprecise region if either the temporal regularity condition is not met or there does not exist the major cluster at the last iteration (refer to Fig.~\ref{fig:overall1}).

\begin{figure}
\centering 
\begin{subfigure}[]{0.45\textwidth}
\centering
\includegraphics[width=2in]{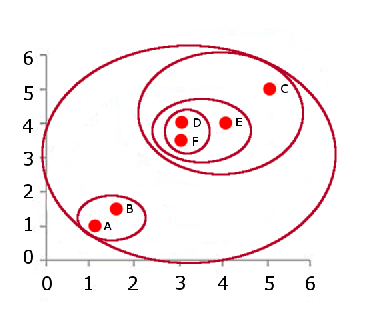}
\caption{Clusters in a two-dimension Euclidean space}
\label{fig:hierarchical_1}
\end{subfigure}
\begin{subfigure}[]{0.45\textwidth}
\centering
\includegraphics[width=2in]{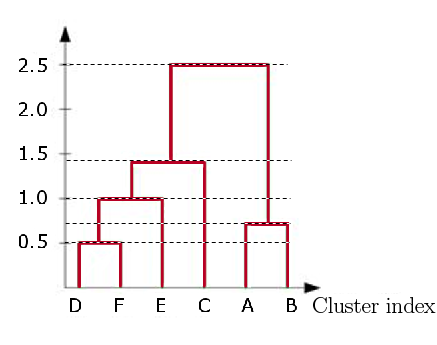}
\caption{The dendrogram representing the hierarchy of clusters}
\label{fig:hierarchical_2}
\end{subfigure}
\caption{An illustration of the single-linkage algorithm, where each red ellipse corresponds to a merged cluster.}
\label{fig:hierarchical}
\end{figure}
\begin{figure*}[t]
\centering 
\includegraphics[width=6.5in]{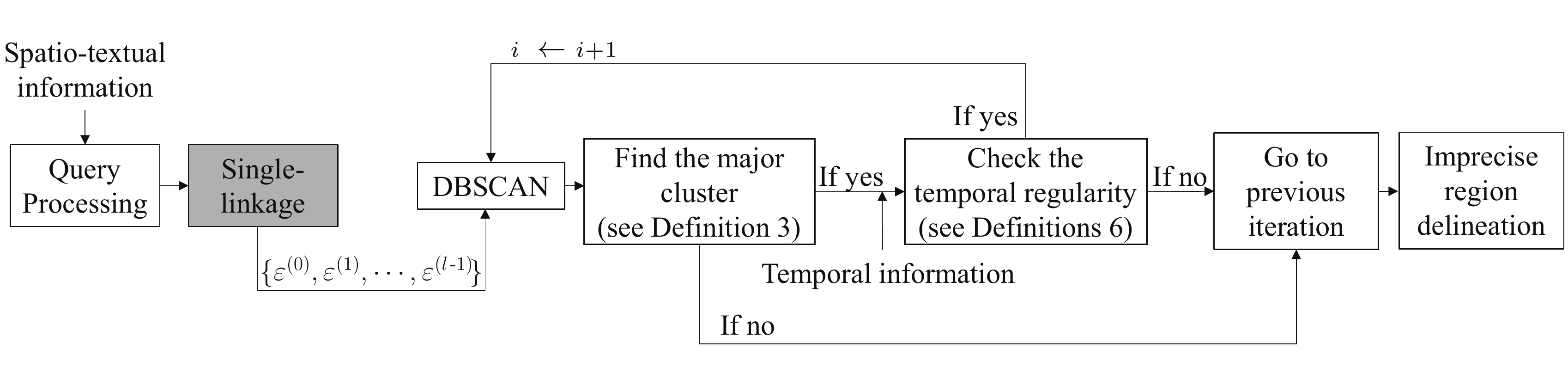}
\caption{The steps of \textsf{DIR-ST$^2$}, including the single-linkage algorithm.}
\label{fig:overall3}
\end{figure*}
\section{Proposed Framework with Hierarchical Clustering}\label{algorithm}
In this section, we describe implementation details of the proposed \textsf{DIR-ST$^2$} framework with an efficient and automated clustering algorithm. More specifically, we aim at finding the radius $\varepsilon$ presented in our problem formulation via {\em hierarchical clustering}, instead of brute-force search over $\varepsilon>0$. The idea behind this application of hierarchical clustering to our framework is due to the fact that in the iterative process in Section~\ref{methodology}, clustering results obtained by DBSCAN change only when there exists the major cluster and the temporal regularity condition is fulfilled. Thus, we focus primarily on investigating such values of $\varepsilon$ that lead to different clustering results.

In our work, we adopt \textit{single-linkage}~\cite{singlelink}, a hierarchical clustering algorithm that computes the closest distances between points so as to generate a dendrogram representing the hierarchy of clusters. A simple example is illustrated in Fig.~\ref{fig:hierarchical}, where six points (namely, A, B, C, D, E, F) are placed in a two-dimensional Euclidean space (see Fig.~\ref{fig:hierarchical_1}). By combining two clusters that contain the closest pair in terms of Euclidean distance when pairwise comparison is made, larger clusters are created in an agglomerative fashion until all points belong to a single cluster (note that a single point can also be regarded as a cluster). The order of merging the clusters in this example is stated as follows:
\begin{enumerate}
\item Cluster \{D\} and Cluster \{F\} are merged.
\item Cluster \{A\} and Cluster \{B\} are merged.
\item Cluster \{D, F\} and Cluster \{E\} are merged.
\item Cluster \{D, F, E\} and Cluster \{C\} are merged.
\item Cluster \{D, F, E, C\} and Cluster \{A, B\} are merged.
\end{enumerate}
This hierarchy of clusters is represented by a dendrogram in Fig.~\ref{fig:hierarchical_2}, where the vertical axis indicates the minimum Euclidean distance between two points that belong to different clusters when two or more clusters are merged (e.g., the minimum Euclidean distance between one point in Cluster \{D, F, E, C\} and another point in Cluster \{A, B\} is given by 2.5). By cutting the dendrogram in a top-down manner for the instance where two clusters are merged (as depicted by dotted lines in Fig.~\ref{fig:hierarchical_2}), we are capable of obtaining the corresponding $l \in \N$ different values of $\varepsilon$, $\{\varepsilon^{(0)},\cdots,\varepsilon^{(l-1)}\}$, that cover all clustering results returned by DBSCAN. Here, $l$ denotes the number of instances where two clusters are merged, which corresponds to the number of dotted lines in Fig.~\ref{fig:hierarchical_2}.

Next, we elaborate on how to integrate the single-link algorithm into our \textsf{DIR-ST$^2$} framework, which is conducted before the iterative process of DBSCAN begins, as illustrated in Fig.~\ref{fig:overall3}. The overall steps of the \textsf{DIR-ST$^2$} framework including hierarchical clustering are summarized in Algorithm 1. One input of the algorithm is the result of the query processing, {\em tweet\_data}, i.e., the tweets relevant to a region of interest consisting of the spatio-temporal information from three fields $lat$, $lon$, and $created\_at$. Another input is a pre-defined threshold $\delta>0$. From the spatial information, we first generate a dendrogram through the function \textsf{SL}, indicating the single-linkage algorithm mentioned above. From the dendrogram, we then use the function \textsf{getEps} to obtain the set of $\varepsilon$'s (i.e., $\{\varepsilon^{(0)},\cdots,\varepsilon^{(l-1)}\}$) sorted in descending order. The radius $\varepsilon^*$ and the set of geo-tagged points corresponding to the desired cluster $C^*$, denoted by $\mathcal{X}_{C^*}$, are initially set to $\varepsilon^{(0)}$ and {\em tweet\_data}, respectively. Then, for the $i$th iteration, we select $\varepsilon^{(i-1)}$ out of the sorted list and calculate the parameter $MinPts$ via the function \textsf{PtsCalc} according to~(\ref{eq:minpts}). The function \textsf{DBSCAN} in line 3 performs DBSCAN and returns the set of clusters, $\mathcal{C}$, the number of clusters, $N$, and the corresponding sets of geo-tagged points, $\{\mathcal{X}_{C_1},\cdots,\mathcal{X}_{C_N}\}$. In lines 4--6, we find the major cluster (refer to Definition 3). Then, in line 10, the function \textsf{EntropyCalc} computes the entropy values of the inter-tweet time intervals for all resulting clusters according to~(\ref{eq:entropy}). As seen in lines 11--12, we next check the temporal regularity condition of the major cluster (refer to Definition 6). The algorithm is terminated when the major cluster does not exist or the temporal regularity condition in~(\ref{eq:regularity}) is not satisfied. The outputs of the algorithm are the radius $\varepsilon^*$ and the set of geo-tagged points corresponding to the major cluster acquired at the second last iteration.

Finally, we delineate the imprecise region as the area covered by the $\varepsilon^*$-neighborhood of every point in the set $\mathcal{X}_{C^*}$.
\begin{table}[t]
\label{algo1}
\renewcommand{\arraystretch}{1.1}
\centering
\begin{tabular}{l}
\hline
\textbf{Algorithm 1} \textsf{DIR-ST$^2$} \\
\hline \textbf{Input}: $tweet\_data, \delta$\\
\textbf{Output}: $\varepsilon^*, \mathcal{X}_{C^{*}}$\\
\textbf{Initialization}: $dendrogram \leftarrow \textsf{SL}(tweet\_data)$;\\
$\{\varepsilon^{(0)},\cdots,\varepsilon^{(l-1)}\} \leftarrow \textsf{getEps}(dendrogram)$;\\
$\varepsilon^* \leftarrow \varepsilon^{(0)}$; $\mathcal{X}_{C^{*}} \leftarrow tweet\_data$;\\
01:\hspace{0.3cm}\textbf{for} $i$ \textbf{from} 0 \textbf{to} $l-1$\\
02:\hspace{0.6cm}$MinPts \leftarrow \textsf{PtsCalc}(\varepsilon^{(i)},tweet\_data)$\\
03:\hspace{0.6cm}$\{\mathcal{C},N,\mathcal{X}_{C_1},\cdots,\mathcal{X}_{C_N}\}$\\
\hspace{1.6cm} $\leftarrow \textsf{DBSCAN}(\varepsilon^{(i)},MinPts,tweet\_data)$\\
04:\hspace{0.6cm}\textbf{for} $j$ \textbf{from} 1 \textbf{to} $N$\\
05:\hspace{1.2cm}\textbf{if} $|\mathcal{X}_{C_j}| = \max\{|\mathcal{X}_{C_1}|,\cdots,|\mathcal{X}_{C_N}|\}$ \textbf{then}\\
06:\hspace{1.8cm}$C_{m} \leftarrow C_j$\\
07:\hspace{0.6cm}\textbf{if} $|\mathcal{X}_{C_m}| \leq |noise|$ \textbf{then}\\
08:\hspace{1.2cm}\textbf{return} ($\varepsilon^*, \mathcal{X}_{C^{*}}$)\\
09:\hspace{0.6cm}\textbf{for} $j$ \textbf{from} 1 \textbf{to} $N$\\
10:\hspace{1.2cm}$H(I_j)\leftarrow \textsf{EntropyCalc}(C_j,tweet\_data)$\\
11:\hspace{0.6cm}\textbf{for} $j$ \textbf{from} 1 \textbf{to} $N$\\
12:\hspace{1.2cm}\textbf{if} $H(I_j)-H(I_{m}) > \delta$ \textbf{and} ${j \neq m}$ \textbf{then}\\
13:\hspace{1.8cm}\textbf{return} ($\varepsilon^*, \mathcal{X}_{C^{*}}$)\\
14:\hspace{0.6cm}$\varepsilon^* \leftarrow \varepsilon^{(i)}$\\
15:\hspace{0.6cm}$\mathcal{X}_{C^{*}} \leftarrow \mathcal{X}_{C_m}$\\

\end{tabular}
\end{table}

\section{Experimental Results}\label{experiment}

In this section, we evaluate the performance of the proposed \textsf{DIR-ST$^2$} framework. For comparison, a state-of-the-art approach for imprecise region delineation using geo-tagged points from social media is employed, where the footprints of geographic regions are discovered by using the OCSVM algorithm~\cite{rw6}. The input parameters of OCSVM are automatically selected by leveraging the spatial distribution of points in precisely-defined regions.

Now, let us turn to describing a parameter setting of our \textsf{DIR-ST$^2$} framework. Since a pre-defined threshold $\delta$ in Definition~\ref{def:6} is used for the temporal regularity condition, it should be set appropriately based on all the entropy values (i.e., $\{H(I_1),\cdots,H(I_N)\}$) obtained from each iteration of Algorithm 1. In our study, we set
\begin{equation}
\delta = \frac{H_{\max}-H_{\min}}{2},
\nonumber
\end{equation}
 where $H_{\max}=\max\{H(I_1),\cdots,H(I_N)\}$ and $H_{\min}=\min\{H(I_1),\cdots,H(I_N)\}$, which guarantees satisfactory performance while leading to an automated algorithm (to be shown in the next subsections).

In the following subsections, we first show experimental results of both the \textsf{DIR-ST$^2$} framework and the state-of-the-art method based on the OCSVM algorithm~\cite{rw6} to delineate {\em precisely-defined} geographic regions, which can be regarded as a ground truth. Then, we examine the performance of the two methods to delineate three {\em imprecise regions}. Finally, we analytically and empirically show the computational complexity of the \textsf{DIR-ST$^2$} framework.

\subsection{Delineation of Precisely-Defined Regions}\label{exp_ground}

To validate the superiority of the proposed \textsf{DIR-ST$^2$} over the OCSVM algorithm, we first perform experiments using the following five precisely-defined regions with their administrative boundaries in the UK: Nottingham, Cambridge, Oxford, Leicester, and Buckingham. In particular, we use the dataset consisting of cities in the UK, which can be obtained from the GADM database.\footnote{Refer to http://www.gadm.org.}

\begin{figure*}
\captionsetup[subfigure]{justification=centering}
\centering 
\begin{subfigure}[b]{.45\textwidth}
\centering 
\includegraphics[height=2.1in]{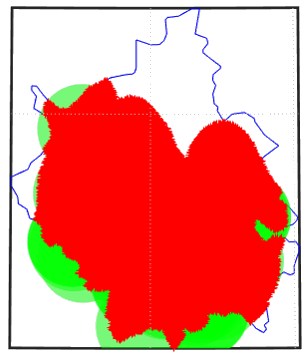}
\caption{Cambridge}
\label{fig:Cambridge}
\end{subfigure}
\begin{subfigure}[b]{.45\textwidth}
\centering 
\includegraphics[height=2.1in]{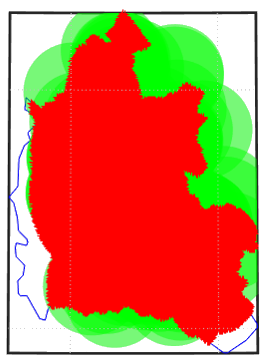}
\caption{Oxford}
\label{fig:Oxford}
\end{subfigure}
\caption{Delineation of two precisely-defined regions using the Monte-Carlo method when \textsf{DIR-ST$^2$} is used. Here, the red area represents the overlapping part between an actual region and a delineated region, the green area is the part of a delineated region that does not overlap with an actual region, and the blue curve represents the administrative boundary.}
\label{fig:figregion}
\end{figure*}

For performance evaluation, the $\mathcal{F}_1$ score is selected due to its popularity in machine learning and statistical analysis. The $\mathcal{F}_1$ score is the harmonic mean of the precision and recall, and is expressed as
\begin{equation}
\mathcal{F}_1 = 2\cdot\frac{Prec \cdot Rec}{Prec + Rec},
\end{equation}
where $Prec$ and $Rec$ indicate the precision and recall, repectively. Let us denote $R$ as the area of an actual region provided by GADM and $\hat{R}$ as the area of a delineated region. Then, the precision and recall can be computed as
\begin{equation}
Prec=\frac{area(R \cap \hat{R})}{area(\hat{R})}
\nonumber
\end{equation} 
\begin{equation}
Rec=\frac{area(R \cap \hat{R})}{area(R)},
\nonumber
\end{equation} 
respectively. Due to the fact that regions tend to be arbitrarily formed, it is difficult to calculate exact areas and their intersections. To overcome this problem, we employ a Monte-Carlo method, which approximates $Prec$ and $Rec$. Specifically, we randomly generate a number of geo-tagged points and then count the number of points inside or outside the corresponding regions. Then, {\em Prec} and {\em Rec} can be approximated as $Prec \simeq \frac{\textrm{number of points in } R  \cap\hat{R}}{\textrm{number of points in }\hat{R}}$ and $Rec \simeq\frac{\textrm{number of points in }R \cap \hat{R}}{\textrm{number of points in }R}$, respectively. As examples, when \textsf{DIR-ST$^2$} framework is used, the clustering results of Cambridge and Oxford are illustrated in Fig.~\ref{fig:figregion}.

\begin{table*}[t]%
    \caption{The performance comparison of \textsf{DIR-ST$^2$} and OCSVM in discovering the footprints of precisely-defined regions}
   \label{tab:performance}
\begin{center}
\begin{tabular}{lcllcllc}
\hline  
                                       & \multicolumn{3}{c}{\textbf{\textsf{DIR-ST$^2$}}}                                                                   & \multicolumn{3}{c}{\textbf{OCSVM}}                                                                     & \textbf{Gain (\%)}
                             \\
\hline  
\multicolumn{1}{c}{\textbf{Region}} & $Prec$         & \multicolumn{1}{c}{$Rec$} & \multicolumn{1}{c}{\textbf{$\mathcal{F}_1$($X$)}} & $Prec$         & \multicolumn{1}{c}{$Rec$} & \multicolumn{1}{c}{\textbf{$\mathcal{F}_1$($Y$)}} & \textbf{$\left(\frac{X-Y}{Y} \times 100\right)$} \\
\hline
\texttt{Nottingham}         & 0.95 & 0.71 & 0.81 & 0.57 & 0.83 & 0.68 & 16\\
\texttt{Cambridge}         & 0.82 & 0.78 & 0.80 & 1 & 0.07 & 0.14 & 66\\
\texttt{Oxford}         & 0.64 & 0.90 & 0.75  & 0.72 & 0.03 & 0.06 & 69\\
\texttt{Leicester}         & 0.58 & 0.97 & 0.72  & 0.45 & 0.62 & 0.52 & 20\\
\texttt{Buckingham}         & 0.50 & 0.99 & 0.66  & 0.50 & 0.05 & 0.09 & 57\\
\hline

\label{tab:result}
\end{tabular}%
\end{center}
\end{table*}

Table~\ref{tab:result} presents the experimental results by discovering the footprints of five precisely-defined regions in the UK. From this table, one can see that \textsf{DIR-ST$^2$} outperforms OCSVM for all cases owing to an increment of $Rec$. This is because OCSVM tends to contain clusters outside the desired region, especially when the region is located near metropolitan cities such as London. The clustering results of Buckingham, Cambridge, and Oxford clearly demonstrate this tendency. On the other hand, by virtue of the temporal information, the \textsf{DIR-ST$^2$} framework enables us to avoid such false clusters. It is shown that \textsf{DIR-ST$^2$} remarkably improves the $\mathcal{F}_1$ score by up to 69\% over the state-of-the-art method. The results demonstrate the effectiveness of incorporating the temporal information into delineating the regions using geo-tagged points from social media.

\subsection{Delineation of Imprecise Regions}

In this subsection, we present delineation of both the \textsf{DIR-ST$^2$} framework and the OCSVM algorithm for the following three imprecise regions in the UK: the Midlands, the South East, and the East Anglia. These are commonly-addressed imprecise regions in the UK in prior studies~\cite{rw3,rw8,rw15}. The delineation performance is demonstrated by showing a different amount of noise points in the set of geo-tagged points for each case.

\begin{figure*}
\captionsetup[subfigure]{justification=centering}
\centering 
\begin{subfigure}[b]{.45\textwidth}
\captionsetup[subfigure]{justification=centering}
\centering
\includegraphics[height=2.5in]{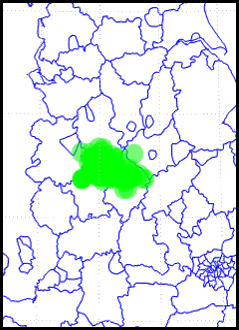}
\caption{The result of \textsf{DIR-ST$^2$}}
\label{fig:fig10a}
\end{subfigure}
\begin{subfigure}[b]{.45\textwidth}
\centering 
\includegraphics[height=2.5in]{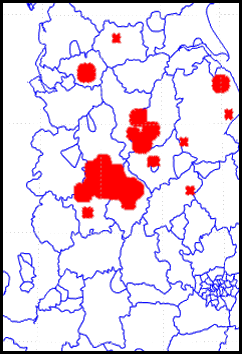}
\caption{The result of OCSVM}
\label{fig:fig10b}
\end{subfigure}
\caption{Delineation of the Midlands.}
\label{fig:fig4}
\end{figure*}

Figure~\ref{fig:fig4} shows the delineation results of the Midlands of the UK. From the figure, the delineated regions that both \textsf{DIR-ST$^2$} and OCSVM return cover the proper zone for the Midlands. However, it seems that the region resulting from the OCSVM algorithm contains many clusters that should be treated as noise (see Fig.~\ref{fig:fig10b}). On the other hand, our method is more robust to noise, i.e., it chooses only one major cluster corresponding to the Midlands and excludes other clusters (see Fig.~\ref{fig:fig10a}). Note that such a trend also takes place with other precisely-defined regions, where the noise points account for 20 to 40\% of the whole geo-tagged points.

\begin{figure*}
\captionsetup[subfigure]{justification=centering}
\centering 
\begin{subfigure}[b]{.45\textwidth}
\centering
\includegraphics[height=2.5in]{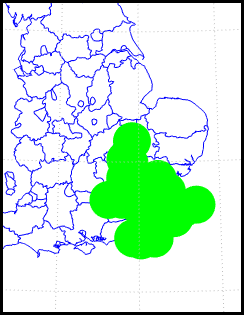}
\caption{The result of \textsf{DIR-ST$^2$}}
\label{fig:fig11a}
\end{subfigure}
\begin{subfigure}[b]{.45\textwidth}
\centering
\includegraphics[height=2.5in]{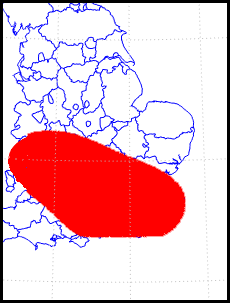}
\caption{The result of OCSVM}
\label{fig:fig11b}
\end{subfigure}
\caption{Delineation of the South East.}
\label{fig:fig5}
\end{figure*}

Next, we present the delineation results of the South East of the UK in Fig.~\ref{fig:fig5}. In this case, since the keyword ``South East" is widely used in many contexts, the number of  region-relevant tweets is huge and  many geo-tagged points from undesired regions are contained. In this experiment, one can clearly see that our framework delineates the region more properly than OCSVM as the delineated region resulting from \textsf{DIR-ST$^2$} is located only in the southeast of the UK (see Fig.~\ref{fig:fig11a}) while the region resulting from OCSVM is spread in the northeast (see Fig.~\ref{fig:fig11b}). This experiment verifies the robustness of the proposed \textsf{DIR-ST$^2$} framework when geo-tagged points in the dataset contain considerably large noise points.

\begin{figure*}
\captionsetup[subfigure]{justification=centering}
\centering 
\begin{subfigure}[b]{.45\textwidth}
\centering
\includegraphics[height=2.5in]{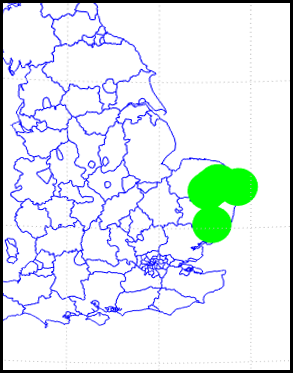}
\caption{The result of \textsf{DIR-ST$^2$}}
\label{fig:fig12a}
\end{subfigure}
\begin{subfigure}[b]{.45\textwidth}
\centering
\includegraphics[height=2.5in]{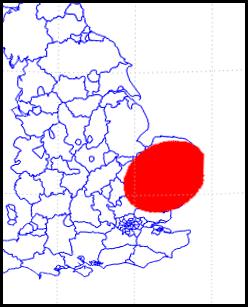}
\caption{The result of OCSVM}
\label{fig:fig12b}
\end{subfigure}
\caption{Delineation of the East Anglia.}
\label{fig:fig6}
\end{figure*}

In addition, we present the delineation results of the East Anglia in Fig.~\ref{fig:fig6}. The results of \textsf{DIR-ST$^2$} and OCSVM are similar to each other due to the concentration of geo-tagged points. Since ``East Anglia" is not a popular keyword, the number of geo-tagged points is relatively small and no noise is found, where only 32 relevant tweets are found. Obviously, the \textsf{DIR-ST$^2$} framework returns the proper zone of the East Anglia.


\subsection{Computational Complexity}

In this subsection, we first analyze the {\em worst-case} computational complexity of the \textsf{DIR-ST$^2$} framework and then empirically show the {\em average} runtime complexity. At the initialization step, we adopt the single-link hierarchical clustering algorithm to find the set of $\varepsilon$'s, each of which indicates the radius of a circle used in the clustering process (see Section~\ref{algorithm}). Since this set has at most $n - 1$ elements, where $n$ is the number of geo-tagged points, the worst case occurs when the points are located away from each other by different distances. Therefore, the number of iterations in Algorithm 1 is at most $n-1$. For each iteration, the DBSCAN algorithm, which has the complexity of $\mathcal{O}(n^2)$~\cite{dbscanrevisit}, dominates the overall complexity. Thus, the worst-case computational complexity of \textsf{DIR-ST$^2$} is bounded by $\mathcal{O}(n^3)$. However, note that such a worst case rarely occurs in most cases since our algorithm is terminated if the temporal regularity condition is not satisfied or the major cluster is not found (see Section~\ref{problemdef}).

\begin{figure}
\centering
\includegraphics[height=1.9in]{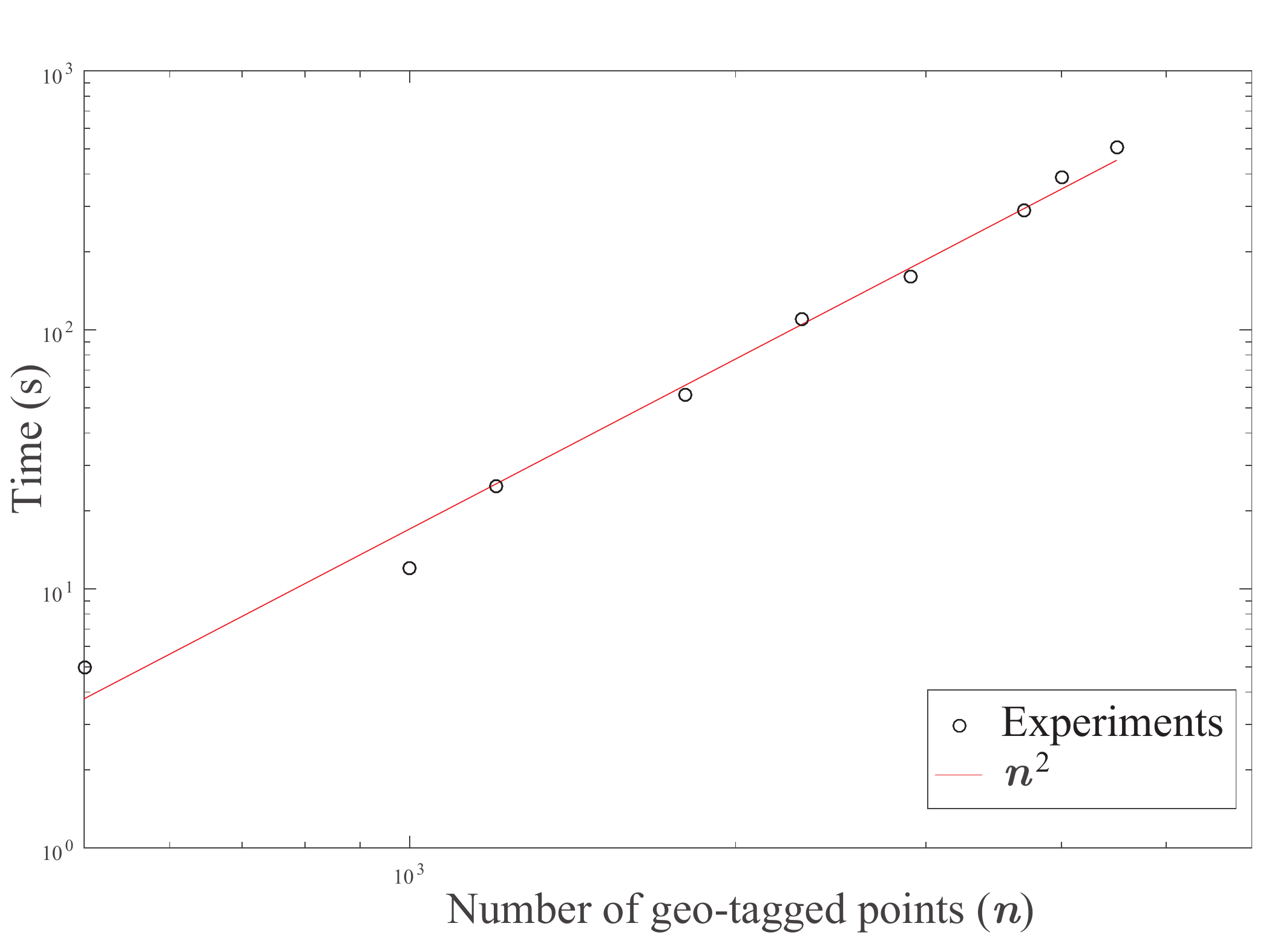}
\caption{The computational complexity of the \textsf{DIR-ST$^2$} framework.}
\label{fig:fig13b}
\end{figure}

Next, we compute the average runtime complexity via experiments when Manchester is delineated. Given different numbers of sampled points from the whole geo-tagged points, i.e., $n\in [500, 4500]$, the computational complexity of the \textsf{DIR-ST$^2$} framework is empirically evaluated. In Fig.~\ref{fig:fig13b}, we illustrate the log-log plot of the runtime complexity in seconds versus the number of geo-tagged points, $n$. An asymptotic curve $n^2$ is also shown in the figure, where it manifests trends consistent with our experimental result. Therefore, it is shown that the average computational complexity of \textsf{DIR-ST$^2$} is approximately given by $\mathcal{O}(n^2)$.

\section{Concluding Remarks}\label{conclusion}

In this paper, we introduced a novel framework, termed \textsf{DIR-ST$^2$}, to automatically and more precisely delineate an imprecise region based on spatio--temporal--textual information on social media. Specifically, our framework was designed in such a way that DBSCAN is iteratively performed by gradually reducing the input parameter $\varepsilon$ and checking the temporal regularity condition with the major cluster for each iteration. In addition, we integrated an efficient $\varepsilon$-search algorithm via hierarchical clustering into the \textsf{DIR-ST$^2$} method. Experimental results showed that from comparison with a ground truth, our \textsf{DIR-ST$^2$} method outperforms the state-of-the-art approach based on OCSVM by a large margin of the recall score due to the significant noise reduction, which leads to an improvement of up to 69\% in terms of $\mathcal{F}_1$ score. Moreover, it was verified that our proposed method returns apparently better delineation of three imprecise regions in the UK, regardless of the amount of noise in a given dataset. It was also shown that for the \textsf{DIR-ST$^2$} framework, the worst case computational complexity is bounded at most by $\mathcal{O}(n^3)$ and the average complexity is approximately given by $\mathcal{O}(n^2)$.

Potential avenues of future research in this area include the complexity reduction of our framework using parallelization.

\section*{Appendix}\label{AppendixA}
 \subsection{Adaptive Parameter Setting of MinPts}

\begin{table}[t]%
    \caption{The $\mathcal{F}_1$ score of the \textsf{DIR-ST$^2$} framework for the two cases where 1) $MinPts$ is adaptively set according to~(\ref{eq:minpts}) and 2) $MinPts$ is found via exhaustive search in such a way that the $\mathcal{F}_1$ score is maximized}
    \label{tab:performance}
\begin{center}
\begin{tabular}{lclclcl}
\hline
\textbf{Region}   & \textbf{$\mathcal{F}_1$ (adaptive)}  & \textbf{$\mathcal{F}_1$ (maximum)} \\
\hline
\texttt{Nottingham}          & 0.81 & 0.85  \\
\texttt{Cambridge}         & 0.80 & 0.82 \\
\texttt{Oxford}          & 0.63 & 0.78 \\
\texttt{Leicester}          & 0.69 & 0.74  \\
\texttt{Buckingham}          & 0.60 & 0.66  \\
\hline

\label{tab:resultminpts}
\end{tabular}%
\end{center}
\end{table}

To verify that the adaptive parameter setting of $MinPts$ in~(\ref{eq:minpts}) guarantees satisfactory delineation performance, we perform experiments using the following five precisely-defined regions (regarded as a ground truth) in the UK: Nottingham, Cambridge, Oxford, Leicester, and Buckingham. We evaluate the performance of our \textsf{DIR-ST$^2$} framework by selecting the $\mathcal{F}_1$ score for the two cases where $MinPts$ is adaptively set according to~(\ref{eq:minpts}) and $MinPts$ is found via exhaustive search over $\{1,\cdots,n\}$ in the sense that the $\mathcal{F}_1$ score is maximized (which is the best we can hope for). Table~\ref{tab:resultminpts} shows that the adaptive setting of $MinPts$ leads to quite comparable performance to the best case for all regions. This signifies that our adaptive parameter setting can be applicable to the $\textsf{DIR-ST$^2$}$ framework so that the entire clustering steps are automated.
\bibliography{IEEEaccessBib} 
\bibliographystyle{IEEEtran}
\end{document}